\documentclass[aip,pop,superscriptaddress,groupedaddress]{revtex4}

 \usepackage{graphicx}
 \usepackage{dcolumn}
 \usepackage{bm}
 \usepackage{amssymb}
 \usepackage{amsmath}
 \usepackage{amsfonts}
 \usepackage{color}
 \usepackage{natbib}
 \usepackage{url}
 \usepackage{hyperref}
 \hypersetup{
   colorlinks,
   linkcolor=blue,
   urlcolor=blue,
 }
\graphicspath{{figs_finales3/}}

\begin{document}

 \title{En route to realistic modeling of the kinetic-MHD interaction between macroscopic modes and fast particles induced by neutral beam injection in tokamaks}
 \author{F.Orain} \affiliation{\emph{CPHT, Ecole Polytechnique, CNRS, 91128 Palaiseau, France}} 
 
\author{G.Brochard} \affiliation{\emph{CPHT, Ecole Polytechnique, CNRS, 91128 Palaiseau, France}}  \affiliation{\emph{CEA, IRFM, 13108 Saint-Paul-lez-Durance, France}} 
\author{T.Nicolas} \affiliation{\emph{CPHT, Ecole Polytechnique, CNRS, 91128 Palaiseau, France}} 
\author{H.L{\"u}tjens} \affiliation{\emph{CPHT, Ecole Polytechnique, CNRS, 91128 Palaiseau, France}} 
\author{X.Garbet} \affiliation{\emph{CEA, IRFM, 13108 Saint-Paul-lez-Durance, France}} 
\author{R.Dumont} \affiliation{\emph{CEA, IRFM, 13108 Saint-Paul-lez-Durance, France}} 
\author{P.Maget} \affiliation{\emph{CEA, IRFM, 13108 Saint-Paul-lez-Durance, France}}


\begin{abstract}
A new model of fast ion source induced by Neutral Beam Injection (NBI) in tokamaks has been implemented in the hybrid kinetic-magnetohydrodynamic code XTOR-K. This source, combined with the collisions also recently implemented, allows for the modeling of realistic slowing-down populations and their interplay with macroscopic modes. This paper describes the Neutral Beam injection and ionization model designed to reproduce experimental configurations and its validation for a typical discharge of the ASDEX Upgrade tokamak. The application to the interaction of NBI-induced fast particles with a kink mode in ASDEX Upgrade demonstrates a resonance between passing particles and the $n=1$ mode. This resonance partially stabilizes the kink mode and induces a radial transport of fast particles. Preliminary results in ITER-like circular geometry show that NBI induces a toroidal torque but has little impact on the kink mode dynamics.
\end{abstract}


 \maketitle

\section{Introduction}

\subsection{Context}
In magnetically-confined fusion devices such as tokamaks and stellarators, understanding and controlling the growth and propagation of macroscopic instabilities in the plasma is a major concern, since these instabilities affect both the performance and the stability of the confinement. A common approach to study these instabilities is the MagnetoHydroDynamic (MHD) framework, which considers the plasma as a hot fluid. These macroscopic modes are therefore called MHD modes.

However, the presence of fast particles in the plasma can not only modify the stability of the MHD modes but also destabilize additional modes (including a wide variety of Alfv\'en Eigenmodes) that would be stable without fast particles. This will notably happen in burning plasmas (e.g. in the ITER tokamak under construction), where energetic $\alpha$ particles of $3.5$~MeV produced by fusion reactions are expected to interact with MHD modes. The mode destabilization or stabilization results from a resonance between the mode frequency and one (or several) of the characteristic fast-particle frequencies: giration, bounce and toroidal precession. In return, the modes can induce a radial transport of the $\alpha$s, which may become unconfined before they have transferred their energy to the rest of the plasma. 

In the particular case of sawtooth instabilities occurring in the core plasma, it is still an open question whether  $\alpha$s will partially stabilize sawteeth in ITER (thus resulting in ``monster sawteeth" likely to trigger Neoclassical Tearing Modes \cite{Chapman_NF10}) and/or whether they will destabilize fishbone modes (whose growth may either prevent the development of monster sawteeth \cite{Gunter_NF99} or result in combined fishbones/sawteeth \cite{Nave_NF91}).

This MHD-fast-particle interplay is a key issue for the performance of ITER and potential future reactors, motivating intensive research on the topic. Since current tokamaks barely produce any fusion reaction, the $\alpha$ dynamics in ITER can be extrapolated from the behaviour of other fast particles present in running devices, in particular the ones injected by heating sources: Ion and Electron Cyclotron Resonant Heating and Neutral Beam Injection (NBI). In operating tokamaks such as JET, ASDEX Upgrade (AUG) and DIII-D, positive ion based NBI systems inject particles at the order of $100$~keV for a total power of $20$~MW in DIII-D and AUG (8 beams of $2.5$~MW each) and $34$~MW in JET (split into 16 beams). In ITER, two negative ion based NB systems of $16.5$~MW each will inject neutrals at $1$~MeV. 

\textbf{This paper focuses on the modeling of the dynamics of the fast ions resulting from the ionization of neutrals injected by NBI and on the way they interact with MHD modes.} Choice is made to consider sawtooth/fishbone modes in this paper; the interaction with other modes such as Alfven Eigenmodes could be the subject of future works based on the same model. The kinetic-MHD framework of XTOR-K used for the study is described in subsec. \ref{method}. The NBI injection and ionization model newly implemented in the XTOR-K code is introduced in sec. \ref{sec2}. The validation of the model in a typical AUG case and the fast particle distribution obtained from the balance between injection and collisions is presented in sec. \ref{sec3}. The resonant interaction between NBI-induced fast ions and a sawtooth-fishbone mode is then described in the AUG case (sec. \ref{sec4}). Last, the impact of one NBI fast ion source is modeled for an ITER-like case in sec. \ref{sec5}. 

\subsection{Hybrid kinetic-MHD model} \label{method}

The modeling of a fast ion beam (and of fast particles in general) does not allow using the MHD framework which represents the plasma as a maxwellian fluid. Therefore, to study fast particle interaction with macroscopic modes, it is necessary to use either a kinetic (eventually gyrokinetic) formalism or an hybrid kinetic-MHD model. It is the choice made in the non-linear hybrid code XTOR-K \cite{Lutjens_JCP22}, which self-consistently couples the two-fluid MHD equations (XTOR-2F part \cite{Lutjens_JCP10}, describing the bulk plasma) with full-f kinetic equations describing the 6D-movement of thermal and/or energetic particles \citep{Leblond_PhD, Brochard_PhD}. The code structure and the complete set of equations can be found in \cite{Brochard_PhD}. A similar coupled approach is used in other hybrid MHD-gyrokinetic (5D) codes such as MEGA \cite{MEGA}, XHMGC \cite{XHMGC} and M3D-K \cite{M3D_K};  however, it was chosen in XTOR-K to keep the full 6D trajectory to model the full gyromotion of the ions. Indeed, the ion larmor radius of an $\alpha$ in ITER (of the order of $10$~cm) will be larger than the resistive layer width proportional to the cubic root of the plasma resistivity ($\eta^{1/3}$), of the order of a millimeter. This means that Finite Larmor Radius effects may play a role in the dynamics. 

XTOR-K was recently upgraded to  self-consistently model the collisions between particles (either of the same species or between different populations). Even though a Langevin approach was first considered \cite{Nicolas_NF17}, a model with binary collisions \cite{Bobylev_PRE} was finally found to be more efficient. Langevin collisions between the particles and the fluid (ion + electron) bulk have also been implemented, such that momentum and energy transfer to the particles is exact; however, the current limit of the model is that no energy and momentum is transferred from the particles to the fluid bulk: this is left to future works. This collision model was validated with typical relaxation tests. 

In parallel to the particle collisions, a new source of fast ions induced by NBI was implemented in XTOR-K. The model, using a similar approach as in Refs. \citep{Schneider, Asunta_CPC15, VENUS}, aims at matching as closely as possible the fast-ion sources induced by NBI in experiments. In particular, close attention was given to reproduce the realistic injection geometry and beam trajectory. Several NBI ports can be considered in parallel. Following the beam trajectories, a realistic ionization model, depending on plasma parameters and beam energy, was implemented, as described below in sec. \ref{sec2}.
Thus, the combination of this NBI module with the collision module allows us to reproduce consistent fast-ion distributions induced by NBI and study their interaction with macroscopic modes and closely compare modeling results with experimental observations. The novelty of this combined approach is that for the first time, the dynamics of a slowing-down fast particle distribution that stems from a realistic NBI source can non-linearly evolve together with MHD modes in a self-consistent manner. This paves the way towards multiple key applications for ITER.

\section{Model of fast-ion source induced by Neutral Beam Injection} \label{sec2}

The kinetic part of XTOR-K is a Particle-in-Cell model, which means that the particle distribution in XTOR-K is reproduced by particle markers. These markers, characterized by a constant weight (each of the markers describes a fixed number of particles), are picked to be representative of the actual particle distribution. At first, in the NBI model, the number of neutral particle markers to inject by NBI at each time step is calculated (part \ref{sec:NB}). 

Second, the trajectories of the neutral particle markers are determined (part \ref{sec:TRAJ}). The NBI injector is considered to be a rectangular source characterized by its central position, length and height. The ``central" beam describes the direction between the central position and a target point, which can be inside or outside the domain. Depending on the central beam and random deviations, each neutral particle marker follows its own trajectory. Along this trajectory, the ionization probability of the neutral marker progressively increases. 

If the neutral marker gets ionized inside the plasma, an ion particle marker is injected in the model at the position of ionization, with a velocity characterized by the direction of the neutral marker trajectory and a norm related to the beam energy (part \ref{sec:INJ}). If the neutral leaves the domain without being ionized, it is considered as lost.

\subsection{Calculation of the number of neutral test particles to inject} \label{sec:NB}
Depending on the injection rate or beam power (both quantities depend on each other and cannot be set independently), a given amount of neutral test particles $\delta N_{markers}$ is injected at each time step $\delta t$. This amount is defined by:
\begin{equation}
\delta N_{real} =\delta N_{markers} \times \rm{Weight}
\end{equation}
The injected power is proportional to the product of the beam energy with the injection rate:
\begin{equation}
P_{NBI} = E_{beam} \times \delta N_{real} / \delta t
\end{equation}
Neutral beams are produced from precursor positive or negative ions partly composed of molecular ions (e.g. in JET and ASDEX Upgrade, positive ion beams contain a fraction of $D_2^+$ and $D_3^+$ in addition to atomic $D^+$). So $\delta N_{real}$ is in fact the number of injected molecules $\delta N_{real, molecules}$. Therefore, we need to consider the energy fraction contained by the different molecules. E.g., for deuterium, we have:
\begin{equation}
\delta N_{real, molecules} = \delta N_{real, D^+} + \delta N_{real, D_2^+} + \delta N_{real, D_3^+} 
\end{equation}
And the number of injected atoms is:
\begin{equation}
\delta N_{real, atoms} = \delta N_{real, D^+} + 2 \delta N_{real, D_2^+} + 3 \delta N_{real, D_3^+}
\end{equation}
\begin{equation}
\delta N_{real, atoms} =  \sum_i ( i \times f_i \times  \delta N_{real, molecules} ) =  \delta N_{real, molecules} \times \sum_i ( i \times f_i  )
\end{equation}
Where we define $f_i$ as the fraction of energy carried by the molecule $D_i^+$: $\delta N_{real, D_i^+}=f_i \times \delta N_{real, molecules}$. \textit{In fine}, the relation between the injected particle markers (representing ion atoms) and the NBI power is:
\begin{equation}
\delta N_{markers} = \frac{P_{NBI}  \times \delta t \times \sum_i ( i \times f_i  )} {E_{beam} \times  \rm{Weight}}
\end{equation}
Among these $\delta N_{markers}$ injected, the fraction of atoms injected at the different energies ($E_{beam}$, $E_{beam}/2$ and $E_{beam}/3$) is the following:
\begin{equation}
 f_{atom,i} = \frac{ i \times f_i} {\sum_i ( i \times f_i )} 
\end{equation}
Note that the user has to be careful with the definition of the energy fraction given in literature. Even though it usually characterises the molecular fraction $f_i$, it sometimes refers to the atomic fraction $f_{atom,i}$. Here it was chosen to use the molecular fraction $f_i$ as input for the code. 

\subsection{Trajectory and ionization of each neutral test particle injected}  \label{sec:TRAJ}
\begin{figure}[h!]
\centering
\includegraphics[width=0.6\textwidth]{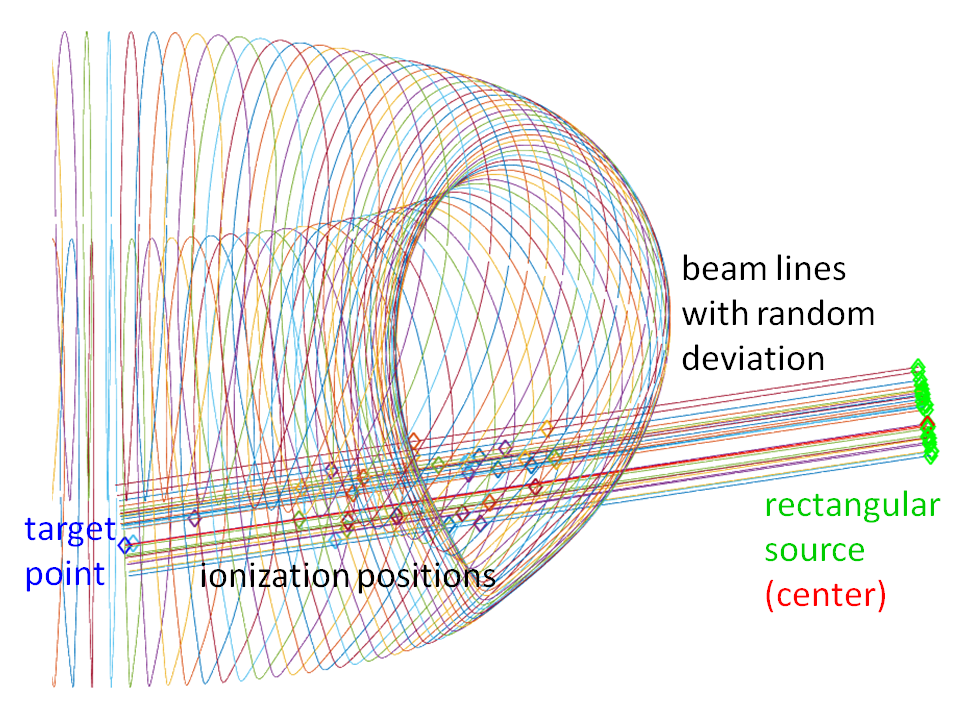}
\caption{Trajectories of several injected neutral particle markers in an ASDEX Upgrade case. The 2D source is plotted in green, with its central point in red. The target point is in blue, and each particle trajectory has a different color. The ionization points of the particles following their own line is represented by the diamonds.} 
\label{fig:fig1}
\end{figure}

Each of the injected neutral particles $\delta N_{markers}$ follow a different trajectory. For each particle, a random starting position is chosen on the 2D rectangular source (characterized by its central position, length and height). Then the particle follows a trajectory towards the target point, corrected with a random spread angle. The line followed by several particle markers is sketched on Fig.\ref{fig:fig1} (ASDEX Upgrade case used for comparison with \cite{Asunta_CPC15}).

Along this line, discretized into infinitesimal pieces $\Delta s$, the electron density $n_e$ and temperature $T_e$ are determined. Depending on these quantities, the cumulative ionization probability $P$ is calculated as follows:
\begin{equation} \label{eq:oneminusP}
1- P(s) = (1-P(s-\Delta s)) \times e^{- \sigma n_e \Delta s}
\end{equation}
where $s$ is the path along the line and $\sigma$ is the stopping cross-section. $\sigma$ is calculated from an analytical fit depending on $n_e$, $T_e$ and $E_{beam}$. This fit, originating from Ref. \cite{Suzuki_PPCF98}, is deduced from atomic processes. Previous cross-section calculations such as in Ref. \cite{Janev_NF89} could also be used.

When a particle follows its trajectory along the line, the cumulative ionization probability $P$ increases. Once $(1-P(s))$ drops below a random threshold $Th$ (with $0<Th<1$), the particle is ionized. In this case, the exact ionization position is recalculated as follows:
\begin{equation}  \label{eq:oneminusP_last}
s = s(n-1) + \Delta s|_{last} = s(n-1) - \frac{1}{n_e \sigma} \times \ln{\Big(\frac{Th}{1-P(s(n-1))}\Big)}
\end{equation}
where $(n-1)$ is the index of the last position where the particle was not ionized yet.

\subsection{Injection of an ion particle marker in the model}  \label{sec:INJ}
At the position where the neutral is ionized, a new ion particle marker is injected in the model. Its velocity is $\vec{v}=\sqrt{2 \times E_{beam}/m_i} \times \vec{u}$, where $\vec{u}$ is the unitary direction of the particle (following the beam direction with random deviation, as explained above). If the neutral particle marker leaves the domain before being ionized, then it is considered as ``shine-through"  and no ion marker particle is injected.

\section{Ionization and collisions in a typical ASDEX Upgrade case} \label{sec3}

\begin{figure}[h!]
\centering
\includegraphics[width=\textwidth]{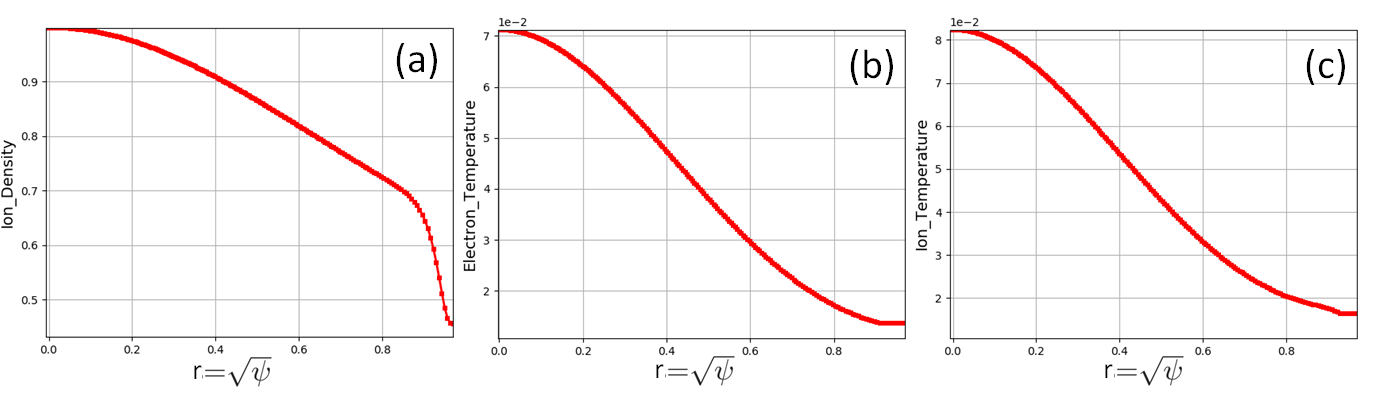}
\caption{Equilibrium radial profiles of (a) ion bulk density and (b) electron and (c) ion bulk temperatures. Density is normalized to the central density $n_{i,0}=7 \times 10^{19}$~m$^{-3}$, and temperatures to the factor $m_p \times V_A^2 / (e \times 10^{-3})$. $V_A$ is the Alfv\'en time, $e$ the electron charge and $m_p$ the proton mass.} 
\label{fig:init_prof}
\end{figure}

The NBI injection and ionization model is validated with a typical AUG case. The geometry of Neutral Beam injectors reproduce the eight injectors used in experiments. The initial profiles of the bulk ion density $n_{i,0}$ as well as electron ($T_{e,0}$) and ion ($T_{i,0}$) temperatures, plotted in Fig. \ref{fig:init_prof}, are chosen to be close to the equilibrium profiles of the AUG discharge \#23076, also used in Ref. \cite{Asunta_CPC15}. This allows for qualitative comparisons with the ionization models of NUBEAM, BBNBI and PENCIL described in Ref. \cite{Asunta_CPC15}. The magnetic field is $B_0=2.67$~T and minor and major radii are $a=0.4$~m and $R_0=1.65$~m. Plasma is approximately taken to have an up-down symmetric D-shape. The computational domain contains only the closed magnetic surfaces. The equilibrium values on the magnetic axis are $n_{i,0}=7 \times 10^{19}$~m$^{-3}$, $T_{e,0}=2.095$~keV and $T_{i,0}=2.42$~keV. Current profile adapted from the $FF'$ equilibrium profile taken from $EQDSK$ calculation, and $P'$ profile calculated from $n_{i,0}$, $T_{e,0}$ and $T_{i,0}$, are computed in the CHEASE equilibrium code \cite{Lutjens_CPC96} to obtain the Grad-Shafranov axisymmetric equilibrium as initial situation. Around 1\% of the bulk is treated kinetically and initialized as a Maxwellian centered in $T_{i,0}$. The initial kinetic distribution allows for the reduction of the noise when particles are progressively injected by the NBI source. The rest of the bulk is treated as a fluid in the MHD module of the XTOR-K code. The resolution is $201$ points in the radial direction, $128$ in the poloidal direction and $32$ in the toroidal direction, where toroidal modes $n=0$ to $5$ are kept in the MHD module. The kinetic and MHD parts self-consistently interact in the code: kinetic pressure tensor is included in the MHD equations and electromagnetic fields calculated from MHD are used for the kinetic movement of particles. 

\subsection{Ionization position for a radial and tangential source}  \label{sec:ioniz}

In a first step, the injection and ionization model is validated alone, without considering collisions and while keeping MHD fields constants. The eight different NBI sources used in AUG are tested and the ionization profiles are compared with the ones described in Ref. \cite{Asunta_CPC15}. As an illustration, we compare the injection of particles of $60$~keV in the radial direction (beam towards the center of the tore) by the Positive Ion-based Neutral Injector (PINI) \#4 and the injection of particles of $93$~keV in the tangential direction (beam oriented close to the toroidal direction) by the PINI \#6. For each NBI, $1920 \times 24~cpu$ markers are injected per step during 1000 time steps of 0.96 Alfv{\'e}n times ($t_A$), which makes $46$ millions of injected markers per NBI during 960 $t_A$. Each marker represents $3.2 \times 10^{12}$ particles. To simplify the validation, only the dominant species is considered ($D^+$) and other energy fractions are neglected. The maximum value of the random spread angle of the beams is set to $20$~mrad ($\sim 1.15^{\circ}$). 

\begin{figure}[h!]
\centering
\includegraphics[width=0.6\textwidth]{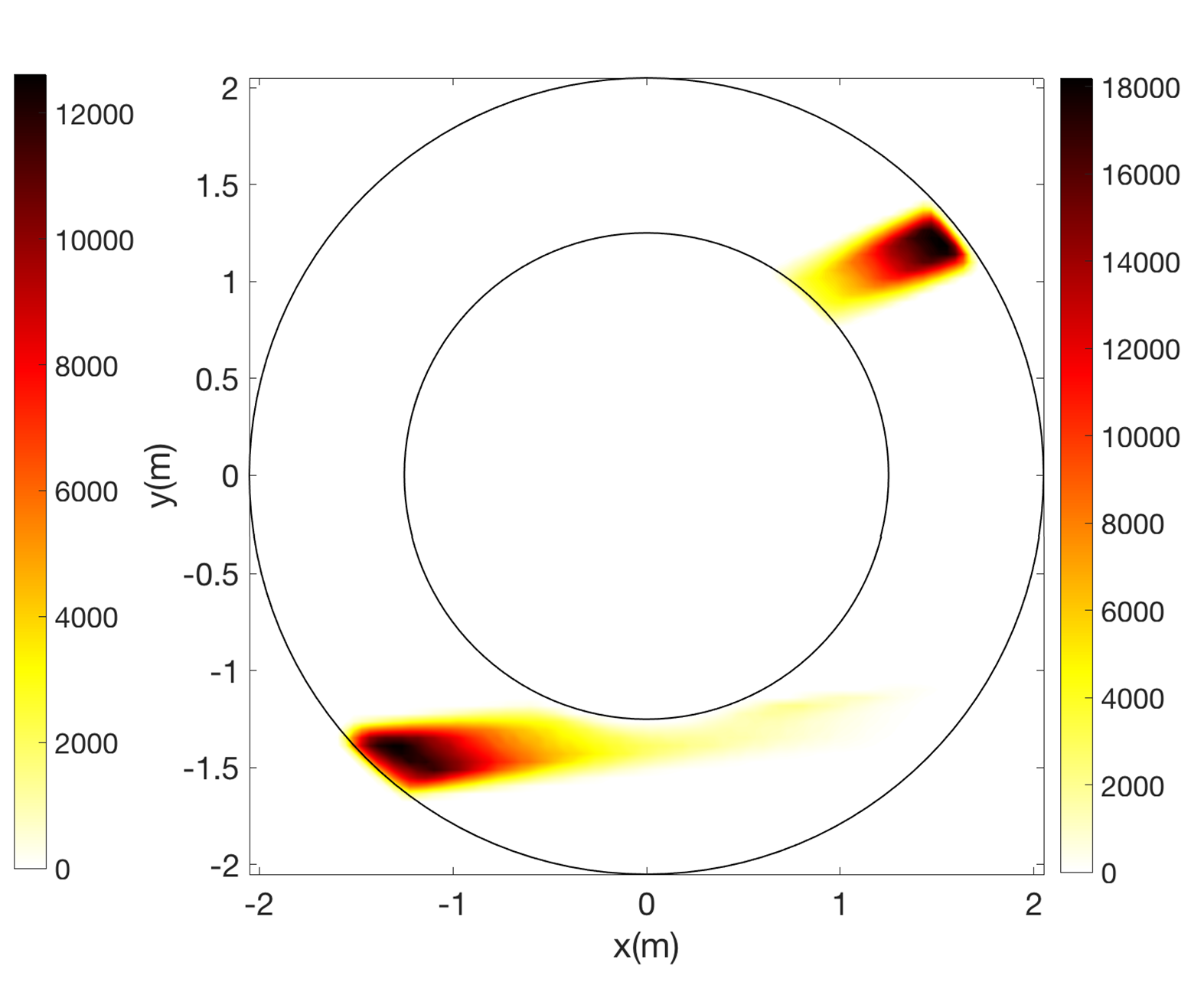}
\caption{Position of ionization of the particles injected by the tangential source (PINI \#6, bottom left) and radial source (PINI \#4, top right). The tokamak is viewed from the top. The left and right colorbars represent the number of injected particles per unit surface integrated on the Z (height) direction, respectively for the tangential and radial sources.}
\label{inj_topview}
\end{figure}

\begin{figure}[h!]
\centering
\includegraphics[width=1.\textwidth]{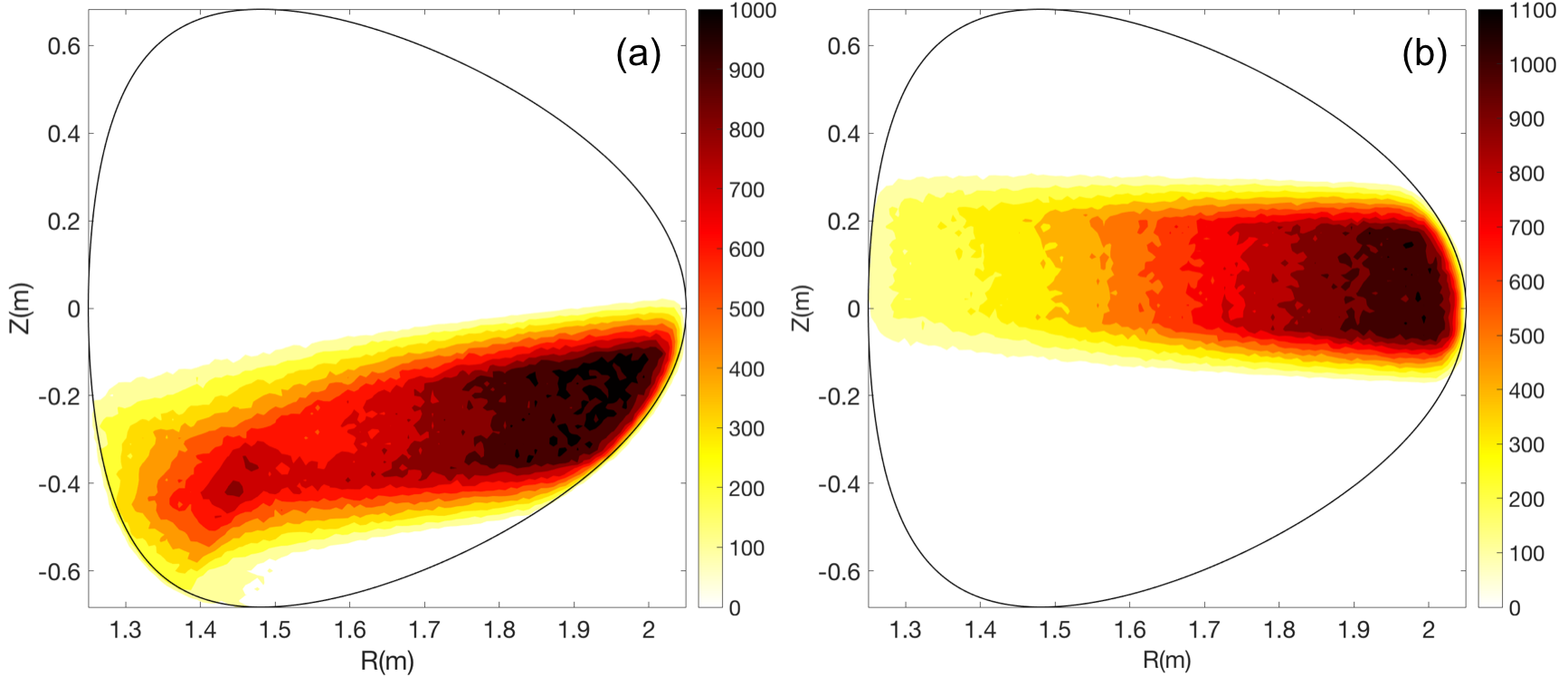}
\caption{Position of ionization of the particles injected by: (a) the tangential source (PINI \#6) and (b): the radial source (PINI \#4). The number of injected particles per unit surface, plotted on a poloidal plane (R,Z), is integrated over the toroidal direction $\varphi$.}
\label{inj_poloidal}
\end{figure}

The position where the injected markers are ionized is projected on the toroidal plane $Z=0$ (tokamak seen from the above) in Fig. \ref{inj_topview} and on the poloidal plane $\varphi=0$ in Fig. \ref{inj_poloidal}. In these plots, each of the directions (x,y) and (R,Z) are discretized for the projection in 100 points and the colors represent the number of markers ionized in each square cell. On the poloidal projection of the toroidal source (Fig. \ref{inj_poloidal}(a)), note that the accumulation of particles below $Z=-0.5$~m is due to the integral over the toroidal direction of the particles ionized after crossing the ``center" ($x=0$~m on Fig. \ref{inj_poloidal}). The amount of particles that leave the domain before being ionized (``shine-through" particles) reach $\sim 9 \%$ for the tangential source and up to $\sim 16 \%$ for the radial source.

The ionization positions resulting from this model qualitatively agrees with the results obtained with other models in Ref. \cite{Asunta_CPC15}. For a more quantitative benchmark, the exact same profiles should be used. Moreover, the approximative up-down symmetric geometry of the computational domain restricted to the closed flux surfaces does not finely reproduce the experimental domain at the edge. This may explain an increased number of particles lost at the edge and thus a larger shine-through fraction in our model compared to Ref. \cite{Asunta_CPC15}. This will be improved in future works using the recent developments to extend the computational domain of XTOR-K up to the wall \cite{Marx}.

\begin{figure}[h!]
\centering
\includegraphics[width=\textwidth]{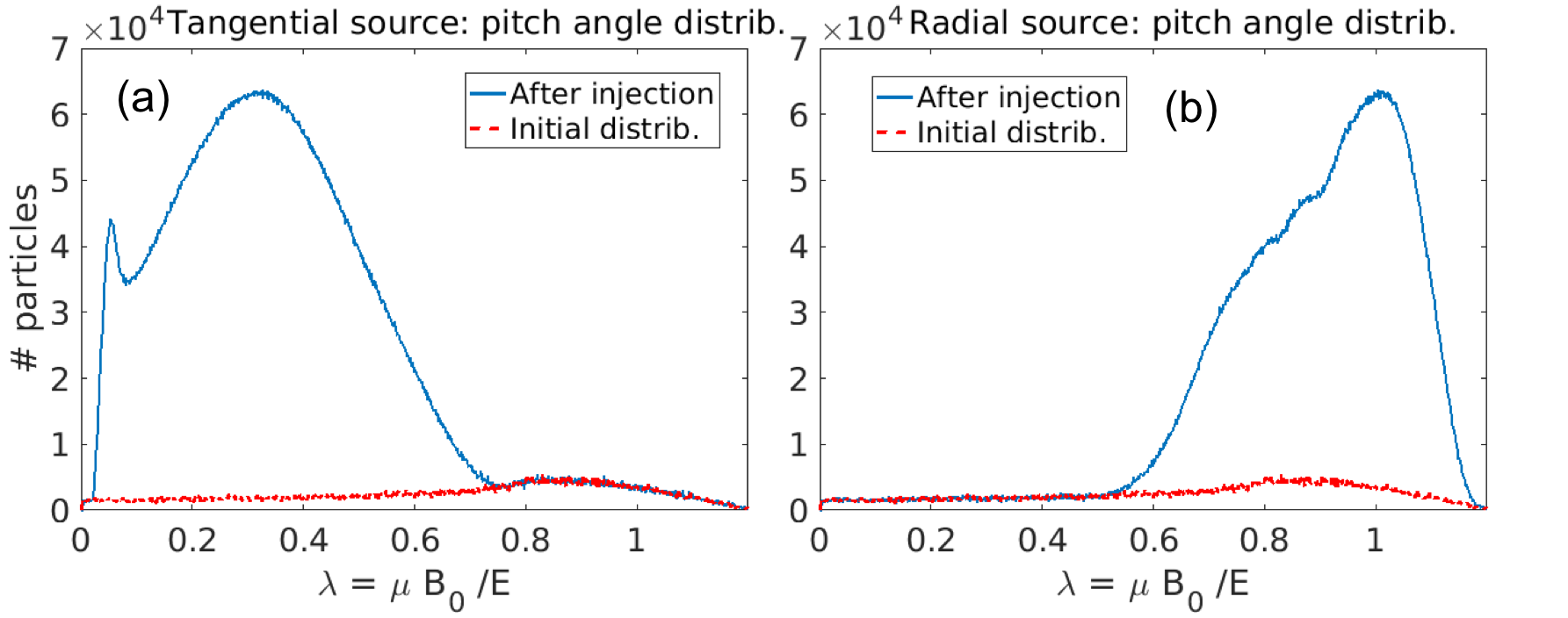}
\caption{Number of injected particles around $q=1$ as a function of their pitch angle $\lambda$, at the initialization of the simulation (red dots) and after the injection (blue line), for the tangential (a) and radial (b) sources respectively. Passing particles are characterized by $\lambda<0.9$ and trapped particles by $\lambda>0.9$.}
\label{rad_tan_lambda}
\end{figure}

\begin{figure}[h!]
\centering
\includegraphics[width=\textwidth]{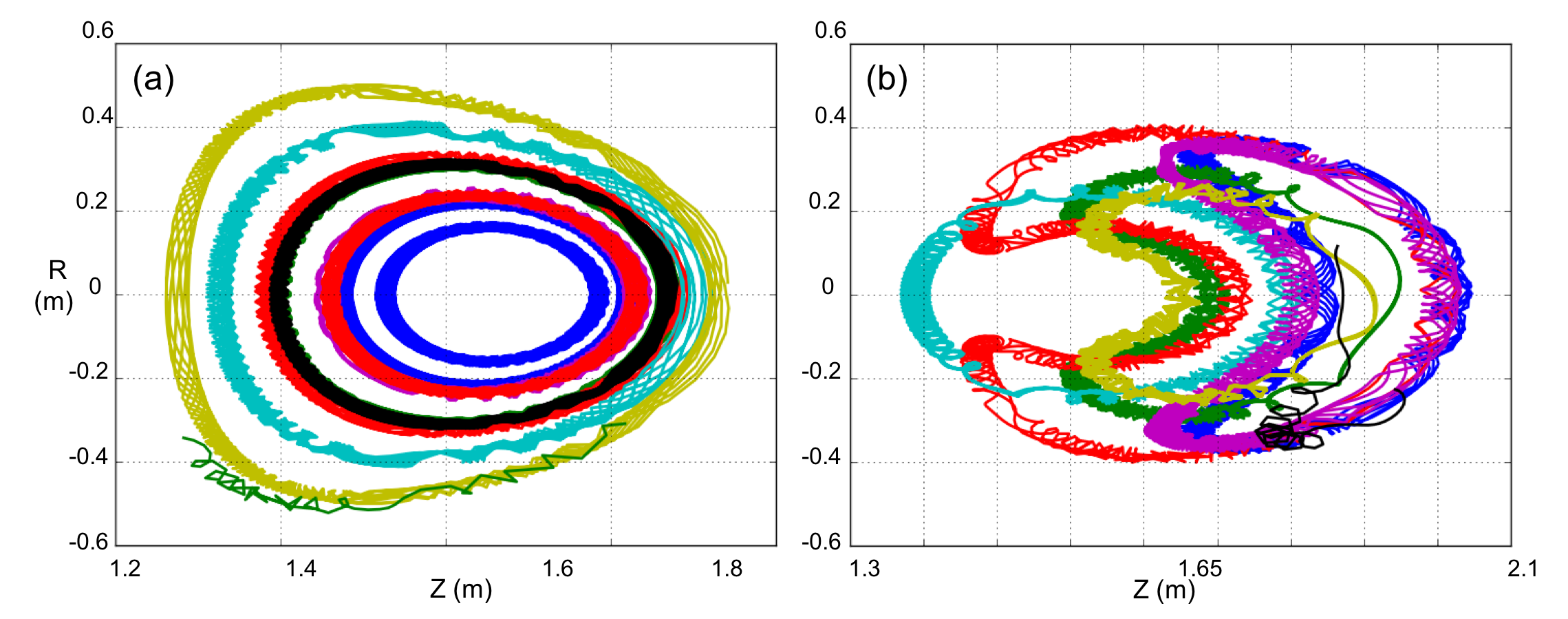}
\caption{Trajectory of a few particles projected in the poloidal plane, highlighting passing particles injected by the tangential source (a) and a majority of trapped particles injected by the radial source (b).}
\label{traj}
\end{figure}

In the phase space, a particle can be characterized by three invariants (without collisions and with constant axisymmetric MHD fields): 1/ its toroidal momentum $P_\varphi$, related to the radial position $r=\sqrt{\psi}$ (where $\psi$ is the normalized poloidal magnetic flux) averaged along the poloidal trajectory; 2/ its energy $E$; 3/ its pitch angle $\lambda$ proportional to the magnetic momentum $\mu$. Fig. \ref{rad_tan_lambda} shows the distribution in pitch angle of the kinetic particles injected around the resonant surface $q=1$ for the tangential source (PINI \#6, left, injection at $E=93$~keV) and the radial source (PINI \#4, right, injection at $E=60$~keV). $q$ stands for the safety factor describing the helicity of the field lines. Particles are passing for $\lambda < 0.9$ and trapped for $\lambda>0.9$. The initial Maxwellian distribution is plotted in red for both cases. In the tangential injection (Fig. \ref{rad_tan_lambda}(a)), the particle parallel velocity $v_{||}$ is large, resulting almost exclusively in passing population. On the contrary, the radial source (Fig. \ref{rad_tan_lambda}(b)) injects particles near-perpendicular of the magnetic field, resulting in a majority of trapped particles. This is confirmed by the plot in Fig \ref{traj} of the trajectory of a few particles injected by the tangential (left) and radial (right) sources.

\begin{figure}[h!]
\centering
\includegraphics[width=\textwidth]{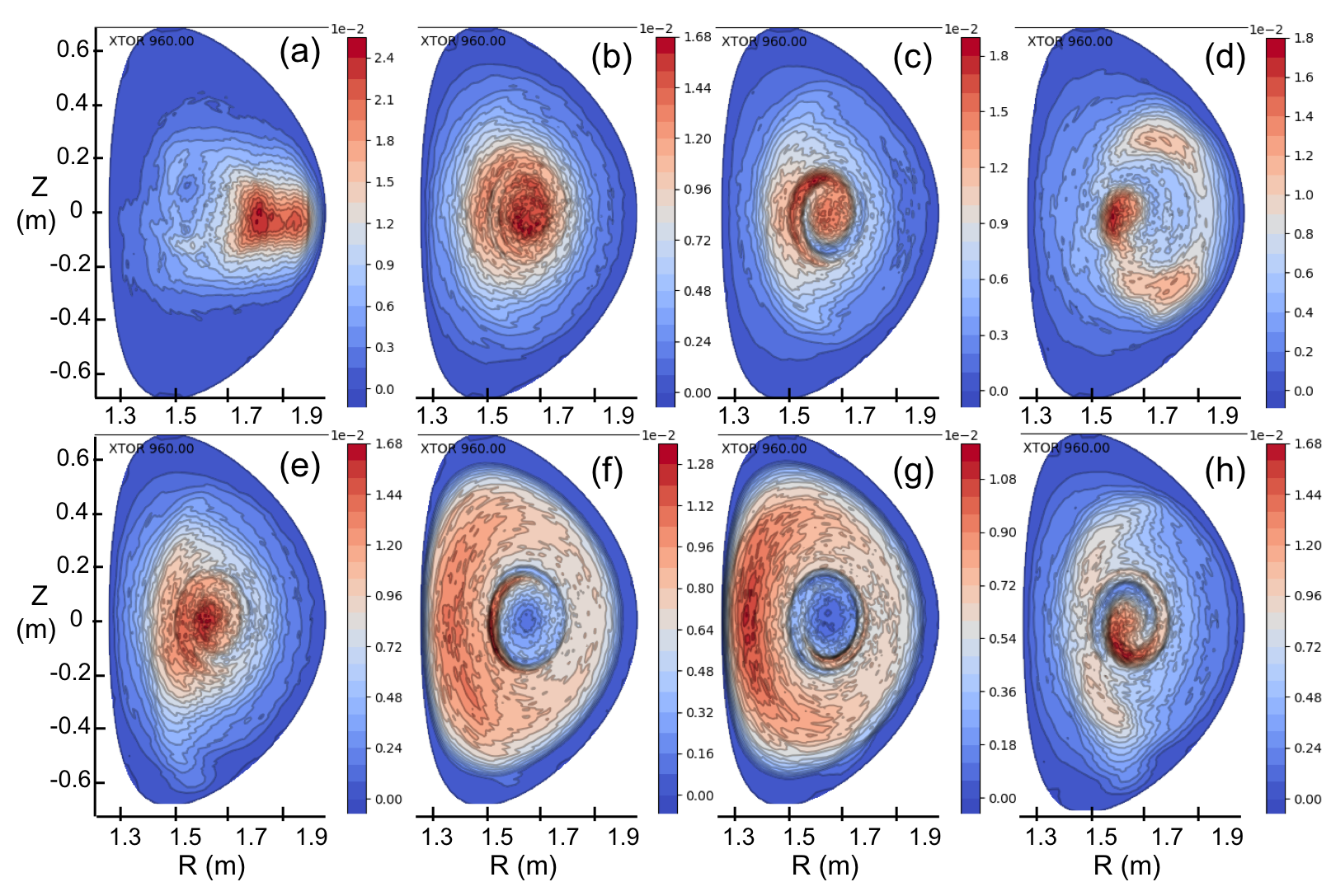}
\caption{(R,Z) profile of the fast particle density in the poloidal plane $\varphi=0$ without collisions after 960 Alfv{\'e}n times of injection by $\#1$ (a) to $\#8$ (h) PINIs. 48000 particle markers are injected per $t_A$.}
\label{inj_nocol}
\end{figure}

The density of ion particles injected by the eight PINI of AUG is plotted in Fig. \ref{inj_nocol}. Each subplot represents the resulting density in the poloidal plane $\varphi=0$ after the continuous injection of 46080 particle markers per unit time step of $0.96~t_A$ (so $48000$ markers per $t_A$). Values are normalized to the central bulk ion density $n_{i,0}$. The injected particles have followed a trajectory imposed by constant electromagnetic fields, since MHD fields are kept constant in this part. After $960~t_A$, the resulting density is partially homogenized over flux surfaces. Note that these plot differ from Fig. \ref{inj_poloidal} representing the position where each particle has been ionized at its injection time.

As observed on Fig. \ref{inj_nocol} - (a) and (d), the injectors \#1 and \#4 induce beams in the radial direction causing a majority of trapped particles. Injectors \#2, \#3, \#5 and \#8 (Fig. \ref{inj_nocol} - b,c, e and h) pointing more in a tangential direction, produce a distribution with a majority of passing particles. As for injectors \#6 and \#7 ((Fig. \ref{inj_nocol} - f and g, also called ``current drive" injectors), they point towards a very tangential direction close to the toroidal direction, resulting in a distribution of mainly passing particles that becomes close to axisymmetric within a few hundreds of Alfv\'en times. 

\subsection{Effect of collisions on injected fast particle distribution}  \label{sec:coll}
In this subsection, the same eight NBI sources are considered but collisions between different particles and between the particles and the fluid (as described in subsec. \ref{method}) are added in the model. In this part, MHD fields are still kept constant to examine the particle distribution function resulting from the balance between sources and collisions. 

This model aims at reproducing fast particle distributions as they are in experiments: $2.5$ MW of power injected per NBI source, with realistic collisions. However, the time scale to thermalize NBI-induced particles is of the order of $10^5-10^6~t_A$ ($\sim 0.05-0.5s)$ for particles of energy $60-100$~keV (AUG), and one order of magnitude more for particles of $1$~MeV as in ITER. Kinetic-MHD simulations on such time scales is too demanding and too resource consuming on nowadays supercomputers. Therefore a simpler approach is to proportionally scale the collision and injection rates to obtain a consistent slowing-down distribution in a reasonable computational time.
 
\begin{figure}[h!]
\centering
\includegraphics[width=\textwidth]{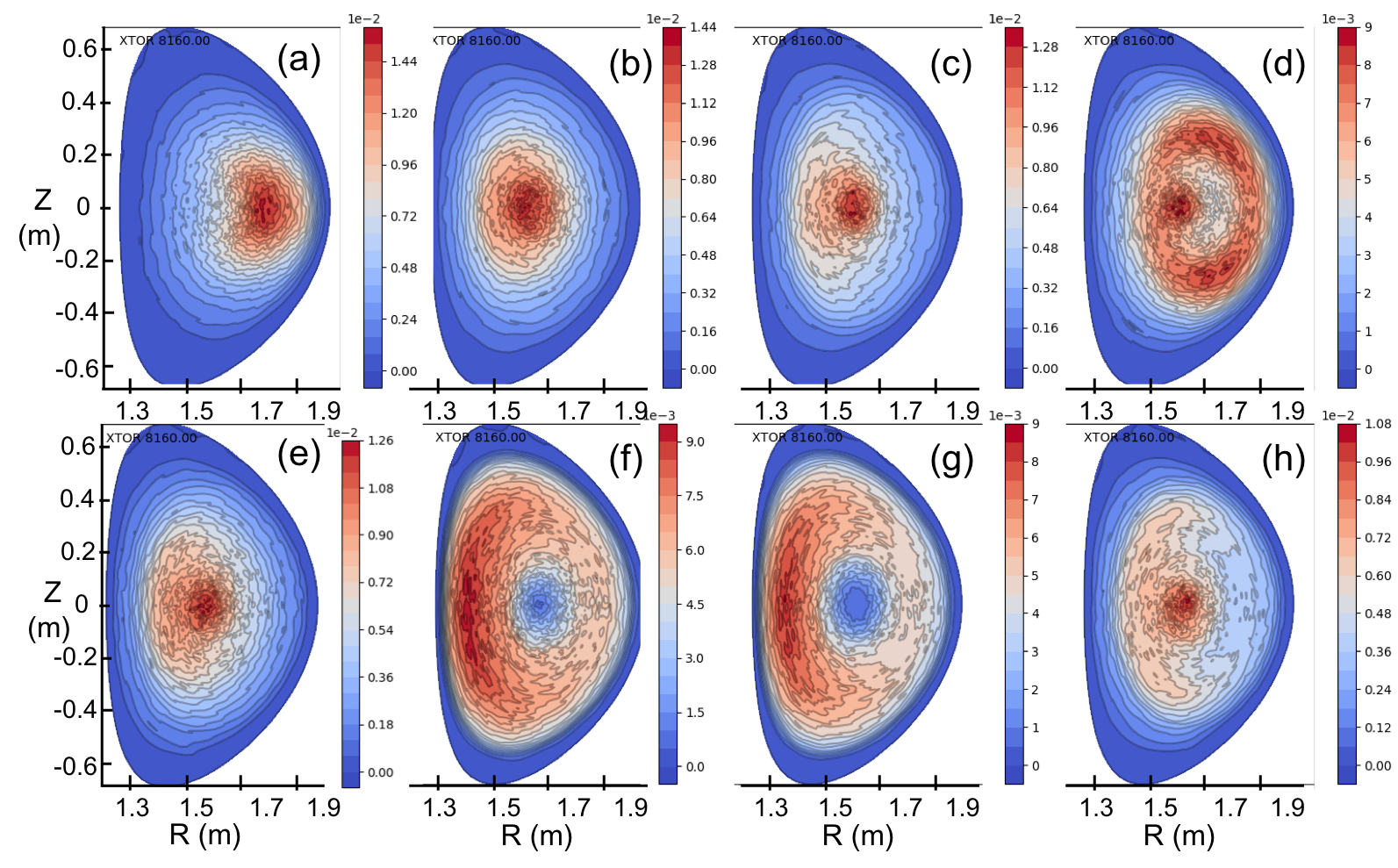}
\caption{(R,Z) profile of the fast particle density in the poloidal plane $\varphi=0$ with both collisions and injection power enhanced by a factor of 10 after 8160 Alfv{\'e}n times of injection by $\#1$ (a) to $\#8$ (h) PINIs. }
\label{inj_col10}
\end{figure}

\begin{figure}[h!]
\centering
\includegraphics[width=\textwidth]{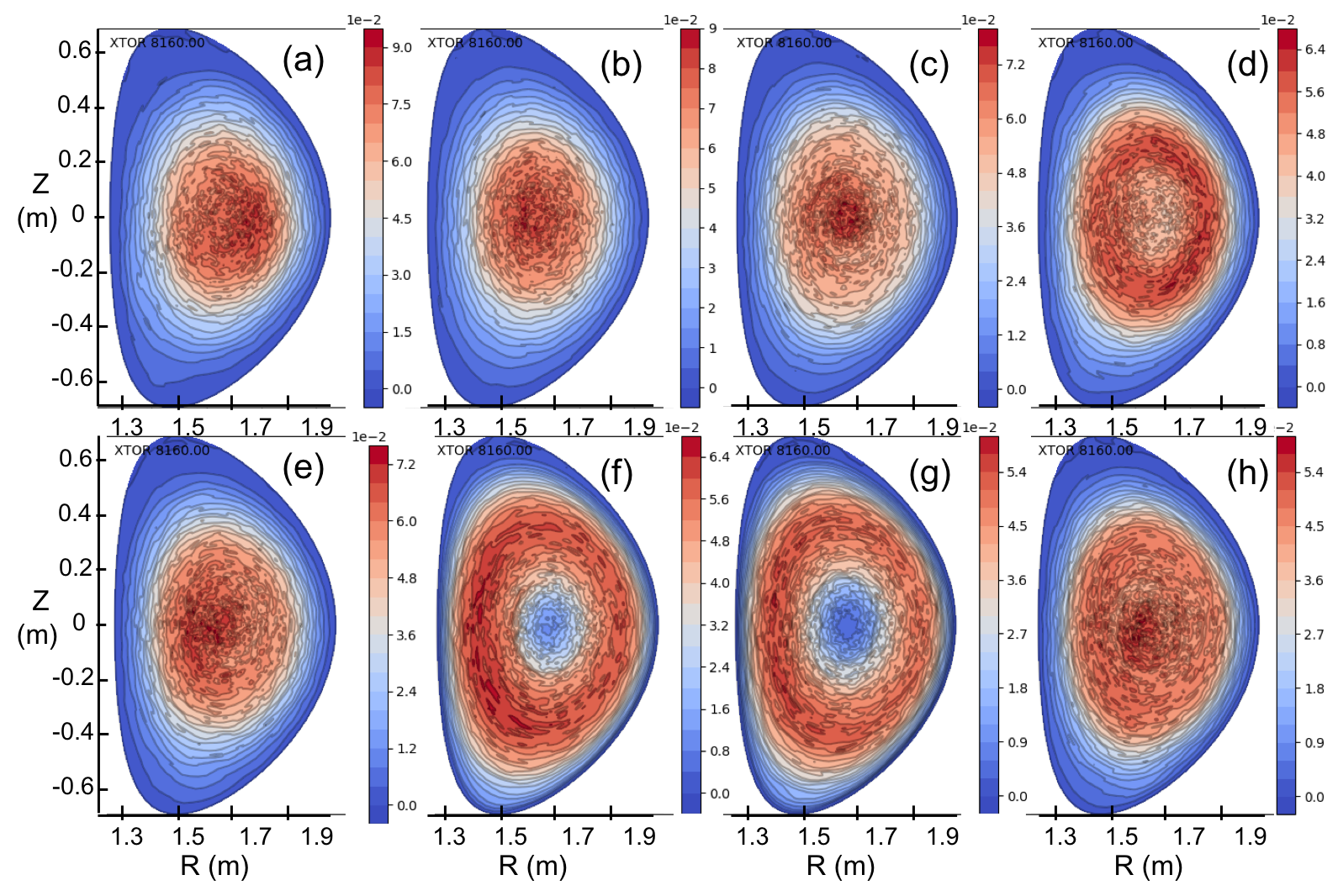}
\caption{(R,Z) profile of the fast particle density in the poloidal plane $\varphi=0$ with both collisions and injection power enhanced by a factor of 100 after 8160 Alfv{\'e}n times of injection by $\#1$ (a) to $\#8$ (h) PINIs.}
\label{inj_col100}
\end{figure}

In Figs. \ref{inj_col10} and \ref{inj_col100}, the fast particle density after 8160 Alfv{\'e}n times is plotted in the poloidal plane $\varphi=0$, in simulations where the nominal injection power ($2.5$~MW) and the collision rate are both increased, either by a factor of 10 (Fig. \ref{inj_col10}) or by a factor of 100 (Fig. \ref{inj_col10}). Each subplot (a-h) corresponds to the injection of particles of $93$~keV with one of the eight PINIs. 
In the case for which collision rate and beam power are both multiplied by 10 (Fig. \ref{inj_col10}), the particle distribution after $8160 t_A$ is a slowing-down that is not yet homogenized over the flux surfaces. However, when collision rate and beam power are multiplied by 100 (Fig. \ref{inj_col100}), the distribution is almost homogeneous on flux surfaces and most of the particles have been thermalized. 

In the next section, we aim at observing resonant interaction between particles and MHD modes. In the latter case (scaling by factor $100$), since most particles are thermalized, too few fast particles are likely to interact with MHD modes. However, in the case scaled by a factor $10$ (Fig. \ref{inj_col10}), the resulting slowing-down distribution fills a wider range of energy and pitch angle values: this is more suitable to observe resonant phenomena. 

\section{Resonance between NBI-induced fast particles and the $n=1$ mode for ASDEX Upgrade-like injection} \label{sec4}

\begin{figure}[h!]
\centering
\includegraphics[width=0.6\textwidth]{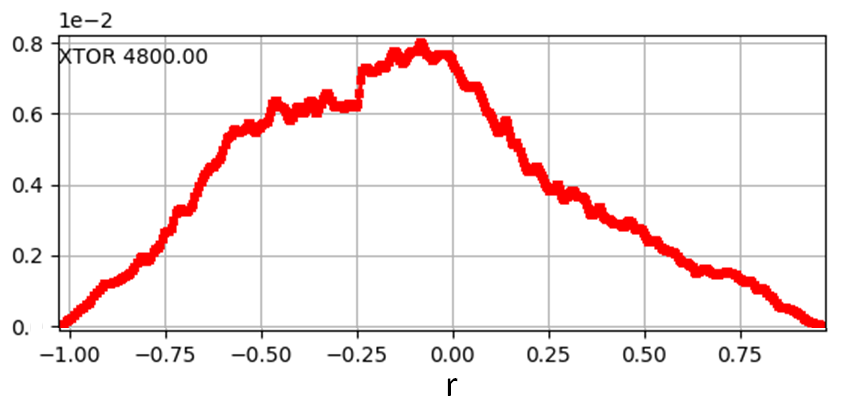}
\caption{Radial profile of the fast particle density at the midplane. The radial direction $r = \pm \sqrt{\psi}$ is normalized to the minor radius.}
\label{nfp_prof}
\end{figure}

The AUG case described in the previous section is used to study resonances between fast particles induced by NBI and MHD modes. In this section, NBI source $\#5$ is used for injection. Collision rate and NBI power are enhanced by a factor 10 to obtain in a reasonable time a slowing-down distribution, as explained above. The reason why NBI source \#5 is chosen here is that a positive fast particle density gradient $\nabla n_{i,fast} >0$ is necessary to obtain the growth of a fishbone mode or the stabilization of a kink mode \citep{Brochard_18, Brochard_20}. Otherwise, if a resonance occurs with $\nabla n_{i,fast} <0$, no resonant transport is induced and thus the resonance cannot be detected. Radial sources \#1 and \#4 were first considered to study trapped resonances but negative $\nabla n_{i,fast} $ in the core (as seen in 2D-profiles in Fig. \ref{inj_col10} (a,d)) makes these configuration unfavorable for the observation of trapped resonances. To study passing resonances, current drive NBI sources (\#6 and \#7, Fig. \ref{inj_col10} (f,g)) also have unfavorable $\nabla n_{i,fast} <0$ in the core. On the contrary, the radial midplane fast particle density profile after injection with NBI source \#5 (Fig. \ref{nfp_prof}) shows a favorable gradient.

For this study, in order to separate the injection dynamics and the mode dynamics, the simulation is run as follows. First, the NBI injection is performed in the kinetic module, coupled with the MHD module in which only the axisymmetric $n=0$ mode alone evolves in time. This allows to reach notably an equilibrium velocity profile modified by the NBI injection, before MHD modes are triggered. 

After the injection during $t= 4800~t_A$, the fast particle pressure fraction (with respect to bulk pressure) has reached $\beta_{fp}=P_{fp}/P_{bulk}=13,9\%$. At this time, the particle injection and collisions are switched off, and MHD modes are added in the simulation. Toroidal modes $n=0$ to $5$ are included. The safety factor in the core was initially set to be below but close to one: $q_0=0.95$, with a quite flat $q$ profile, until $q=1$ for the radial position $r=0.4$ (normalized to the minor radius). This initial $q$ profile is plotted in black in Fig. \ref{energy_q} (b).

\subsection{Impact of fast particles on the $n=1$ mode}  \label{sec:n1mode}

\begin{figure}[h!]
\centering
\includegraphics[width=\textwidth]{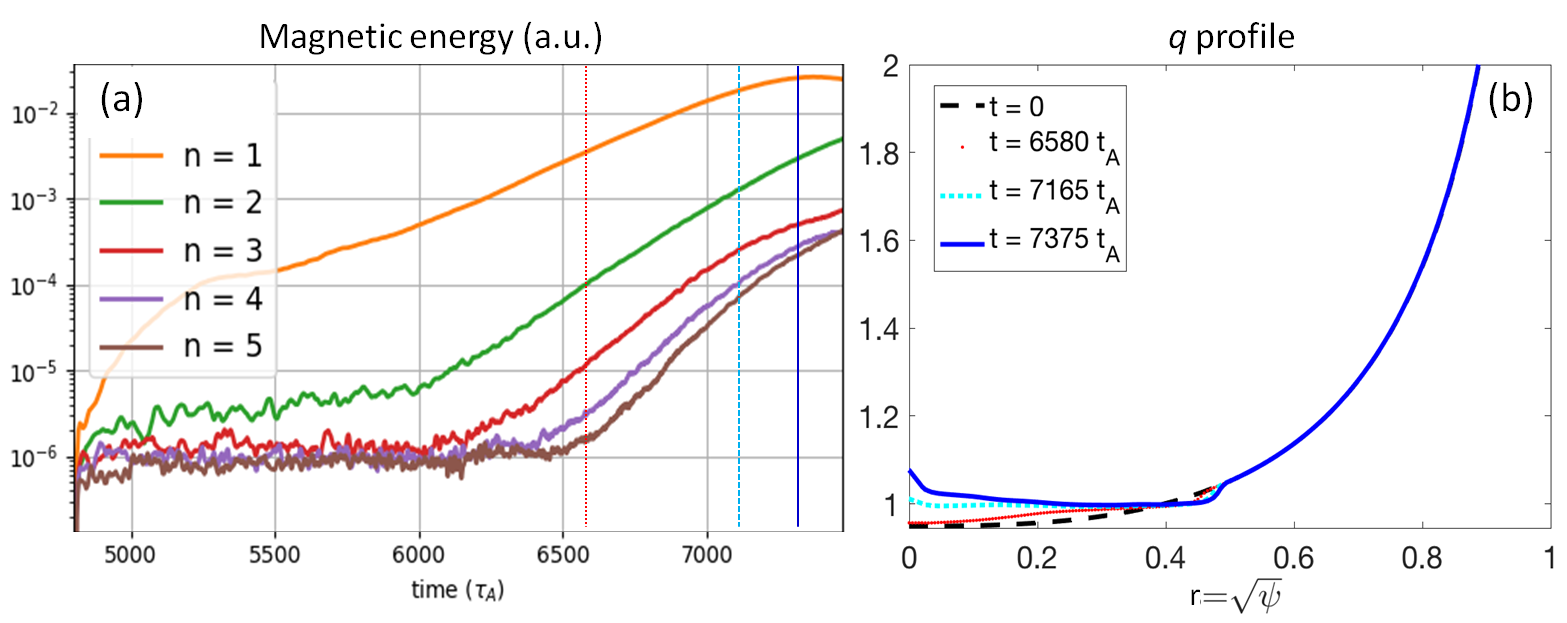}
\caption{(a) Time evolution of the magnetic energy of the toroidal modes $n=1-5$ (arbitrary units). The lines indicate the time slots chosen in (b) with same colors. (b) Radial profile of the safety factor $q$ at the times $t/t_A=0$ (black dashed line, same profile for $t/t_A=4800$ when $n \ne 0$ modes are included), $t/t_A=6580$ (linear phase, red small dots), $7165$ (beginning of non-linear phase, cyan big dots) and $7375$ (blue line).}
\label{energy_q}
\end{figure}

From the reference time $t=4800~t_A$ on, MHD modes are included in the simulation. The unstable $q$ profile in the core very quickly triggers an internal kink mode $(n=1, m=1)$, with $n$ and $m$ the toroidal and poloidal mode numbers. Fig. \ref{energy_q} (a) shows that  the magnetic energy of the $n=1$ mode grows exponentially (linear phase) from $t=5500~t_A$. The harmonics $n=2-5$ follow from $t=6000~t_A$. The kink mode makes the $q$ profile relax in the core until $q$ becomes 1 in the whole core (up to $r=0.4$, initial position of $q=1)$ at $t=7165~t_A$ (cyan dots in Fig. \ref{energy_q} (b)). Later on ($t=7375~t_A$, blue line in Fig. \ref{energy_q} (b)), the $q$ profile even becomes greater than 1 in the core.

\begin{figure}[h!]
\centering
\includegraphics[width=\textwidth]{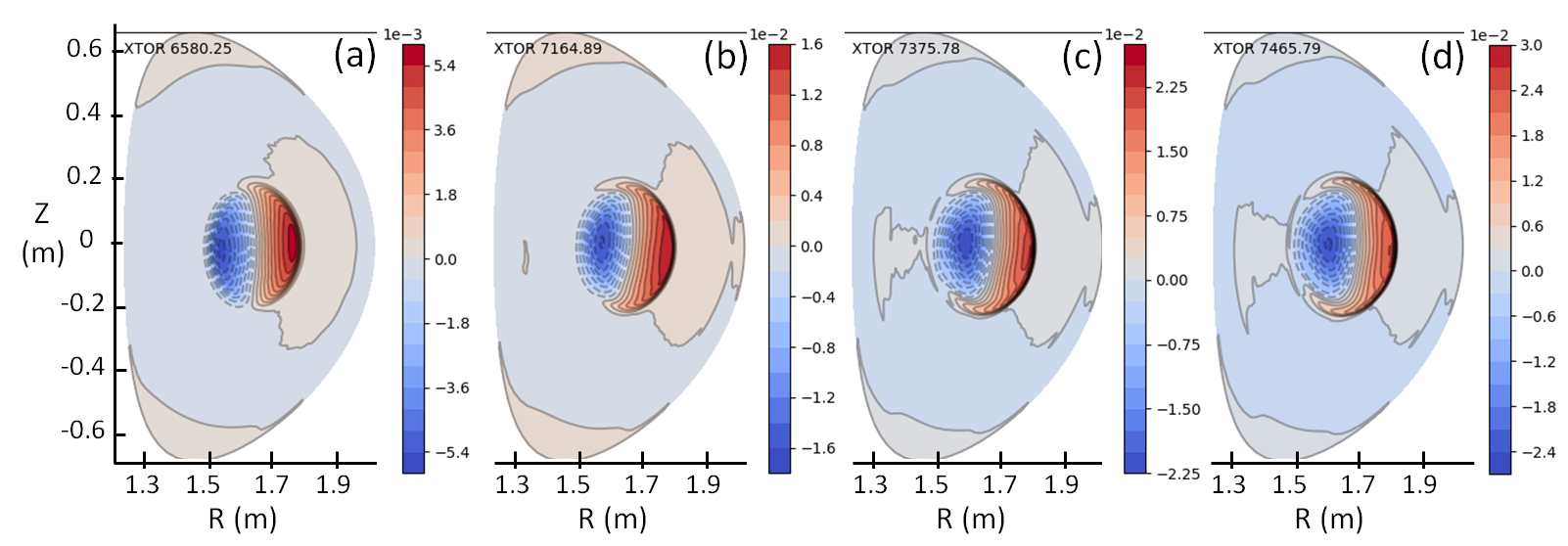}
\caption{Poloidal profile of the pressure perturbation at $\varphi=0$ at the times $t/t_A=6580$ (a), $7165$ (b) and $7375$ (c) and $7465$ (d).}
\label{deltaP}
\end{figure}

Since the diamagnetic rotation is not included in the MHD model (a 1-fluid MHD model is used in this case), the $n=1$ mode does not rotate without fast particle injection. However, the NBI source induces the slow rotation of the kink mode at a frequency $\omega \approx 5 \times 10^{-4} / t_A$. This can be seen on the movement of the bulk pressure perturbation, Fig. \ref{deltaP}, slightly rotated in the clockwise direction between $t=6580$ and $7165~t_A$ (a-b).

\begin{figure}[h!]
\centering
\includegraphics[width=0.5\textwidth]{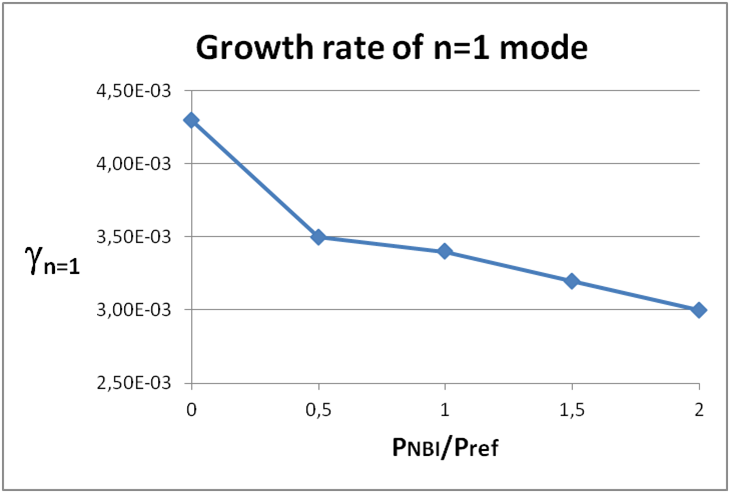}
\caption{Growth rates of the magnetic energy of the $n=1$ mode as a function of the NBI power applied. Growth rates are normalized to the Alfv\'en time.}
\label{gammas}
\end{figure}

In order to check the effect of the NBI source on the $n=1$ mode, a scan in NBI power is performed. The growth rate of the magnetic energy of the $n=1$ mode depending on the NBI power is plotted in Fig. \ref{gammas}. The ``reference power" $P_{ref}$ is the case previously described: both realistic power and collisions are multiplied by 10. Note that for all the applied powers, the same reference collision rate is applied. Fig. \ref{gammas} shows that the growth rate of the $n=1$ mode is reduced when the NBI power is increased, and thus when the fast particle fraction $\beta_{fp}$ (ranging from $0\%$ without NBI up to $20\%$ for $P_{NBI} = 2 \times P_{ref}$) is increased accordingly. Therefore in this case, fast particles have a stabilizing effect on the $n=1$ kink mode, but they do not seem to destabilize fishbone modes in this range of $\beta_{fp}$. It is likely that the destabilization of fishbones (``fishbone branch" with growth rates increasing with $\beta_{fp}$, as observed in Ref. \cite{Brochard_20}) requires unrealistically large fast particle pressure for this plasma configuration.

\subsection{Evidence of resonance between fast particles and the $n=1$ mode}  \label{sec:resonance}
\begin{figure}[h!]
\centering
\includegraphics[width=\textwidth]{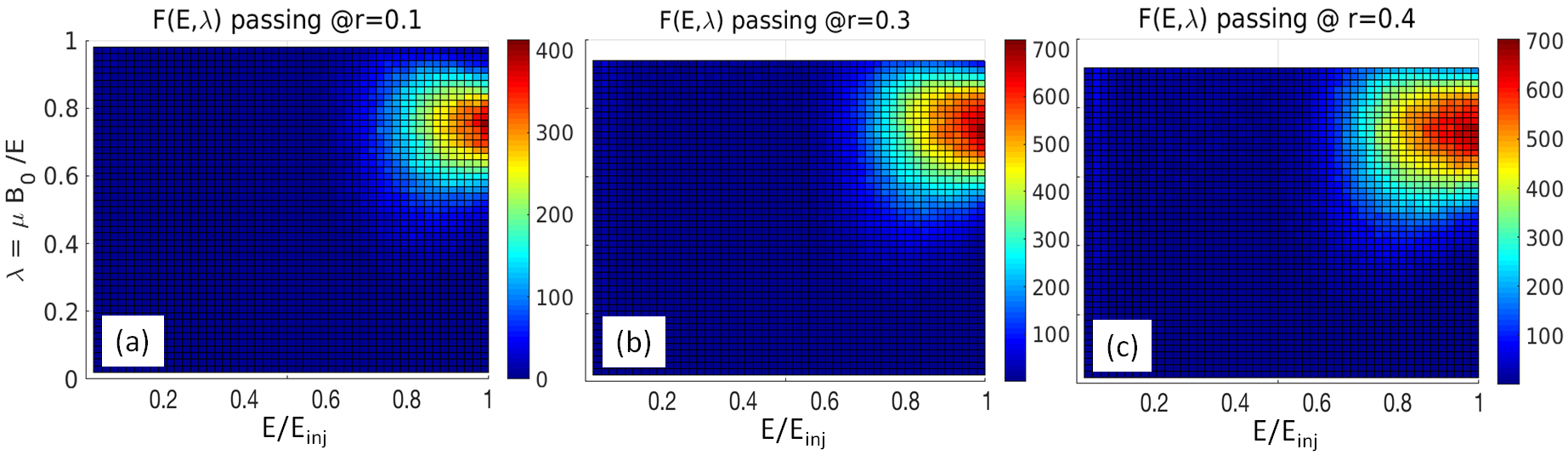}
\caption{Phase space diagrams: distribution function $F(E,\lambda)$ of passing particles at the reference time (when the injection is stopped and when MHD modes $n \ne 0$ are added in simulation), as a function of the energy E (normalized to injection energy $= 93$~keV) and the pitch angle $\lambda$ around three different radial positions: $r=\sqrt{\psi}=0.1$ (a), $0.3$ (b) and $0.4$ (c).}
\label{Fsuppl}
\end{figure}

We wish to understand if the stabilizing effect of fast particles on the $n=1$ kink mode comes from a resonant effect. For this purpose, we study how the passing particle distribution evolves as compared to the ``reference time" when the injection is stopped and when MHD modes are included. 
 In Fig. \ref{Fsuppl}, the passing particle distribution function at the reference time, depending on the particle energy $E$ and pitch angle $\lambda$, is plotted around three different radial positions: inside $q=1$ for $r \approx 0.1$ (left) and $r \approx 0.3$ (middle) and around $q=1$ ($r \approx 0.4$, right). The energy $E$ is normalized to the injection energy ($E_{inj}=93$~keV). It shows that most of the particles injected by the NBI source \#5 and slowed down by collisions lie in the area where $E > 0.6 \times E_{inj}$ and $\lambda > 0.4$.

\begin{figure}[h!]
\centering
\includegraphics[width=\textwidth]{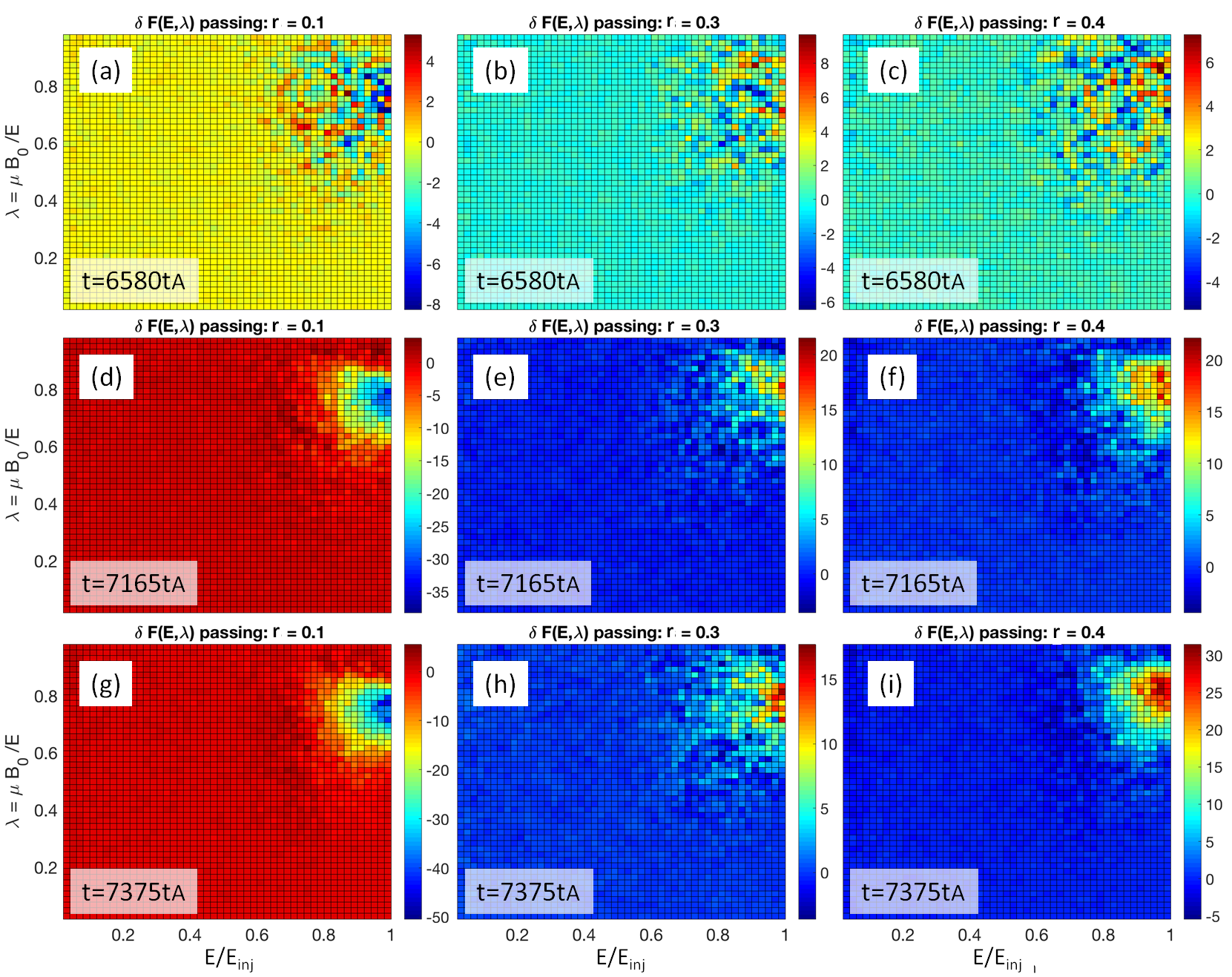}
\caption{Phase space diagrams: variation $\delta F$ of the distribution function of passing particles as compared to the reference distribution, chosen at the time when the injection is stopped and when MHD modes $n \ne 0$ are added in simulation. $\delta F$ is plotted as a function of energy E (normalized to injection energy $= 93$~keV) and pitch angle $\lambda$, for three different times: $t=6580~t_A$ (a-c), $t=7165~t_A$ (d-f) and $t=7375~t_A$ (g-i), around three different radial positions: $r=\sqrt{\psi}=0.1$ (a,d,f), $0.3$ (b,e,h) and $0.4$ (c,f,i). The colorbars represent the variation in percentage.}
\label{deltaF}
\end{figure}

 In Fig. \ref{deltaF}, the variation $\delta F(E,\lambda)$ of the passing distribution function as compared to the reference time, is plotted for three different times at the same three radial positions: $r \approx 0.1,~0.2$ and $0.3$. During the early linear phase ($t=6580~t_A$, Fig. \ref{deltaF} a-c), no clear accumulation pattern is distinguishable in the phase space. However, in the later linear phase ($t=7165~t_A$, d-f) and the non-linear phase ($t=7375~t_A$, g-i), in the region ($E>0.8, \lambda>0.6$), a strong reduction of the passing particle distribution is observed far inside the core (up to 50\% of reduction locally around $r=0.1$, Fig. \ref{deltaF} (d) and (g)). Closer to $q=1$, in the same $(E,\lambda)$ region, an accumulation of particles is remarkable around $r=0.3$ (the number of particle rises by $15-20\%$, Fig. \ref{deltaF} (e) and (h)) and around $r=0.4$ (increase up to $30\%$, Fig. \ref{deltaF} (g) and (i)). This highlights a clear transport of particles not only from the core to the initial position of $q=1$ ($r=0.4$) but also in the phase space $(E,\lambda)$.

We notice that the area in the phase space ($E, \lambda$) presenting a depletion (for $r=0.1$) or an accumulation of particles (for $r=0.3$ and $0.4$) (Fig. \ref{deltaF}) covers most of the region where the reference distribution function $F$ is concentrated (Fig. \ref{Fsuppl}). Therefore, the question arises whether the particle displacement in the physical space and in the phase space is only a radial transport of the whole distribution caused by the kink mode, or whether it is induced by a resonant interaction between the particles and the $n=1$ kink mode. To answer this question, we need to compare the kink mode frequency with the particle frequencies. Let us first define the characteristic particle frequencies $\Omega_{1-3}$, as defined in the angle-action formalism \citep{Brochard_18, Brochard_PhD}.
$\Omega_1$ is the cyclotron frequency:
\begin{equation}
\Omega_1 = \omega_c
\end{equation}
$\Omega_2$ and $\Omega_3$ are respectively the poloidal and toroidal transit frequencies. $\Omega_2$ is the bounce frequency:
\begin{equation}
\Omega_2 = \omega_b
\end{equation}
As for $\Omega_3$, its expression is more complex: 
\begin{equation}
\Omega_3 = \omega_d + q(\bar{r})\epsilon_b\omega_b
\end{equation}
where $\omega_d$ is the precession frequency, $\epsilon_b$ equals 1 for passing particles and 0 for trapped particles, and $\bar{r}$ is the reference flux surface of the particle.

The resonant condition between a particle and an MHD mode is:
\begin{equation}
\omega -n_1 \Omega_1 - n_2 \Omega_2 - n_3 \Omega_3 = 0
\end{equation}
with $\omega$ the MHD mode frequency and $n_1,~n_2,~n_3$ integers. Since $\Omega_1$ is several orders of magnitude larger than $\Omega_2$,  $\Omega_3$, and MHD frequencies, it cannot induce resonant effects: $n_1=0$. Therefore, as explained in Refs. \citep{Brochard_18, Brochard_20}, the only resonance available with the $n=m=1$ internal kink mode is the precessional resonance characterized by $n_1 = n_2 = 0, n_3 =1$:
\begin{equation}
\omega - \Omega_3 =0
\end{equation}
And for passing particles, the passing resonance with the $n=1$ mode is obtained for $n_1 =0, n_2= -1, n_3=1$:
\begin{equation} \label{eq:reson}
\omega + \Omega_2 - \Omega_3 = 0
\end{equation}

\begin{figure}[h!]
\centering
\includegraphics[width=\textwidth]{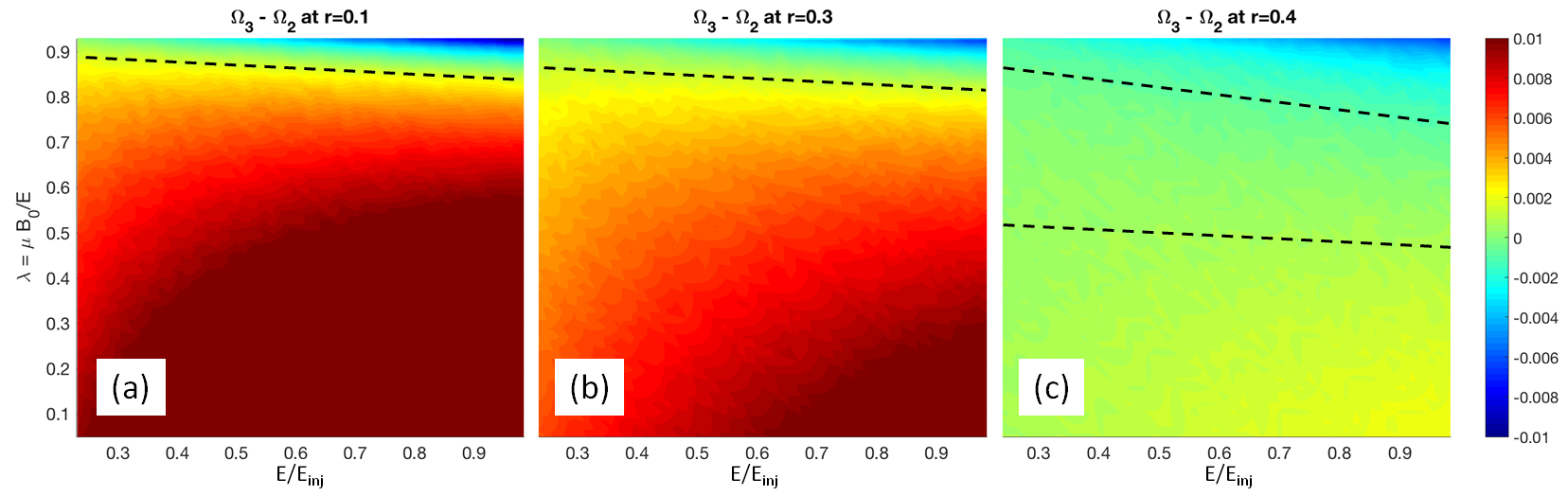}
\caption{Difference between the toroidal and the poloidal transit frequencies of the particles located around three different radial positions: $r=\sqrt{\psi}=0.1$ (a), $0.3$ (b) and $0.4$ (c), as a function of their energy $E$ (normalized to the injection energy) and their pitch angle $\lambda$. In (a-b), the black dashed lines represent the optimal resonant region where $\Omega_3 - \Omega_2 = \omega \approx 5 \times 10^{-4} / t_A$. In (c), the resonant region lies between the two lines.}
\label{omegas}
\end{figure}

In order to check if the passing resonance (equation \ref{eq:reson}) occurs in our simulations, diagnostic particles covering the full ($E, \lambda$) diagram are shot to extract the values of $\Omega_2$ and $\Omega_3$ from the time evolution of their poloidal and toroidal trajectories. The value of $\Omega_3 - \Omega_2$ is plotted for counter-passing particles around $r=0.1, 0.3$ and $0.4$ in Fig. \ref{omegas} (a,b,c respectively). The kink mode frequency $\omega = 5 \times 10^{-4}$ (normalized to the Alfv{\'e}n frequency) is indeed close to the value of $\Omega_3 - \Omega_2$ for counter-passing particles, in the region of the phase space diagram where particles are accumulated in Fig. \ref{deltaF}. This shows that a counter-passing resonance indeed occurs between the NBI-induced fast particles and the kink mode. Note that no resonance occurs with co-passing particles (with opposite sign of $\omega_b$), since their frequencies $\Omega_3 - \Omega_2$ do not match the kink mode frequency $\omega$. Therefore, the counter-passing resonance is found to be responsible for the partial stabilization of the kink mode when the fast particle pressure is increased.

\section{NBI impact in an ITER-like case} \label{sec5}

In this section, an ITER-like case is considered: the toroidal magnetic field is $B_t=5.3~$T, and the minor and major radii are $r=2$~m and $R_0=6.2$~m. Typical density and temperature values are used for the bulk plasma, with core values: $n_{i,0}=1 \times 10^{20}$~m$^-3$, $T_{i,0}=T_{e,0}=20$~keV. For simplicity and for comparisons with the linear model for kink-fishbone stability described in Refs. \citep{Brochard_18, Brochard_20}, a circular cross-section is chosen. The resolutions taken in radial, poloidal and toroidal directions respectively are: 201, 128 and 32 points. The current profile is set such that the safety factor is flat in the core, with central value $q_0=0.96$ and $q=1$ for $r=\sqrt{\psi}=0.35$.

\begin{figure}[h!]
\centering
\includegraphics[width=0.75\textwidth]{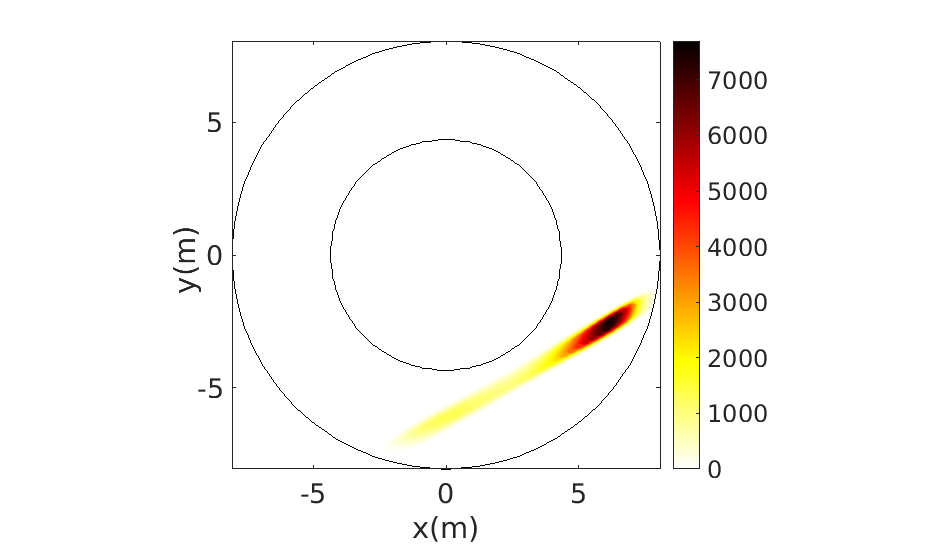}
\caption{Position of ionization of the neutrals injected by NBI \#2. The tokamak is seen from the above. The number of particle is integrated over unit surface ($x$ and $y$ directions are subdivided into 100 pieces for the integration).}
\label{ITER_injection}
\end{figure}

The bulk plasma is considered as a fluid (MHD) except $4\%$ of the bulk that is treated kinetically and initialized as a Maxwellian centered in $T_{i,0}$. From this initial situation, a beam is injected with the ITER Neutral Beam Injector \#2, whose geometry is taken as defined in Refs. \citep{ITER_NBI, ITER_doc}. Particles of energy $E=1$~MeV are injected by the beam. No half or third energy fraction is considered here for simplicity. For the same reason of computational feasibility as in the previous AUG case, nominal NBI power in ITER ($16.5$~MW) is enhanced in simulation by a factor of $500$ and realistic collision rate by a factor $1000$. Since the particles injected in ITER are ten times more energetic than in the AUG case, such a large scaling factor is necessary to slow them down in a reasonable computational time. The two-times larger scaling factor used for the collision rate with respect to the injection rate enables to quickly reach a typical fast particle fraction of $\beta_{fp}=13.4\%$. Particles are continuously injected during $8000~t_A$. The ionization position of the fast ions issued from the beam is plotted in Fig. \ref{ITER_injection}. The tokamak is viewed from the top in this plot. 

\begin{figure}[h!]
\centering
\includegraphics[width=\textwidth]{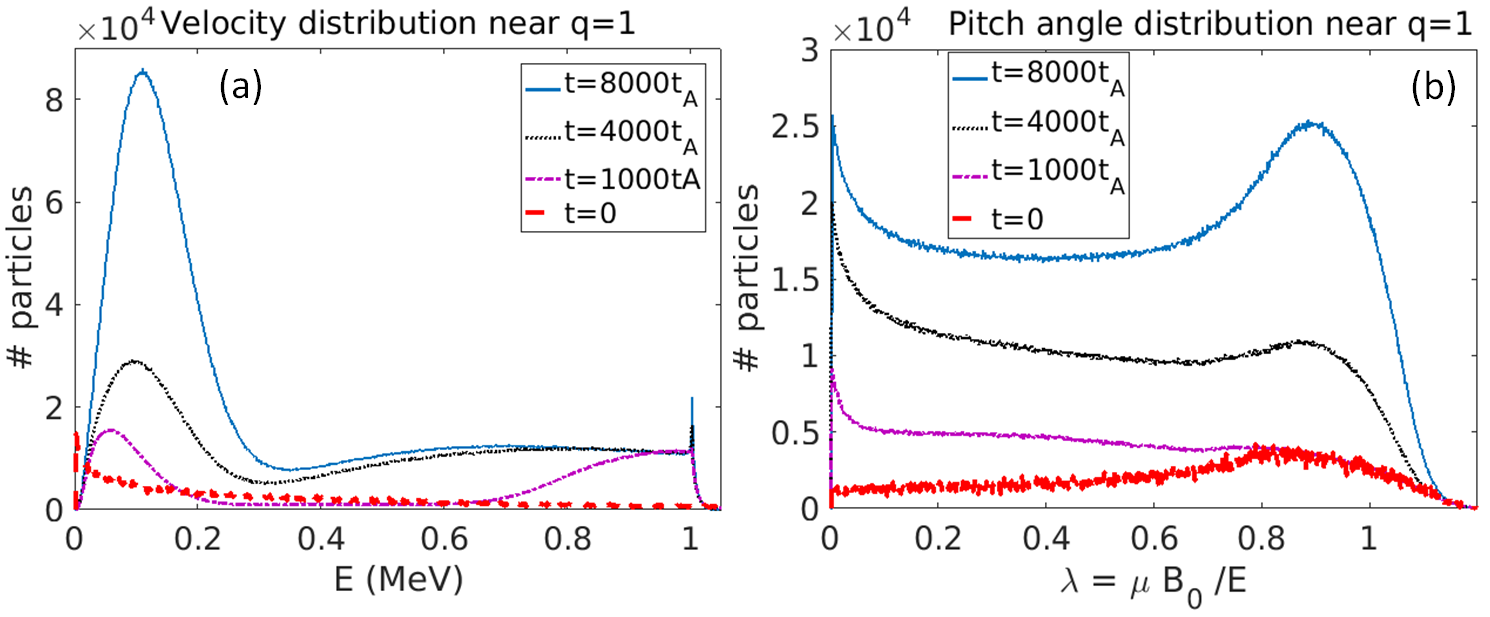}
\caption{Distribution of kinetic particles as a function of their energy $E$ in MeV (a) and or their pitch angle $\lambda$ (b), at the following times: $t=0$ (initial Maxwellian distribution, red dashed line), $t=1000~t_A$ (purple dash-dots), $t=4000~t_A$ (black dots) and $t=8000~t_A$ (blue line). Particles are injected at the energy $E=1$~MeV and with a pitch-angle distribution peaked in $0$. Particles are passing below $\lambda=0.88$ and trapped above.}
\label{ITER_Elambda}
\end{figure}

At the time when the injection is stopped ($t_{1}=8000~t_A$), $16.3$ millions of particle markers have been injected in total, generating a slowing-down fast particle distribution with a fast particle pressure fraction of $\beta_{fp}=13.4\%$. The evolution in time of the distribution is provided in Fig. \ref{ITER_Elambda}, as a function of the particle energy $E$ (a) and pitch angle $\lambda$ (b). The very tangential direction of ITER NBI induces a population of very passing particles characterized by small $\lambda$ values, peaked in $0$. Due to collisions, particles with initial energy $E=1$~MeV and small pitch angle cool down, resulting in a tail in energy $E<1$~MeV and pitch angle $\lambda>0$ that progressively extends. The initial bulk kinetically-treated (in red) collides with fast particles; at the same time, a portion of fast particles thermalizes, resulting in a Maxwellian population heated to $\sim 100$~keV in addition to the slowing down tail. Note that at present, the rest of the bulk plasma that is treated as fluid (MHD) does not gain energy from collisions. This will be the subject of future works. 

\begin{figure}[h!]
\centering
\includegraphics[width=0.5\textwidth]{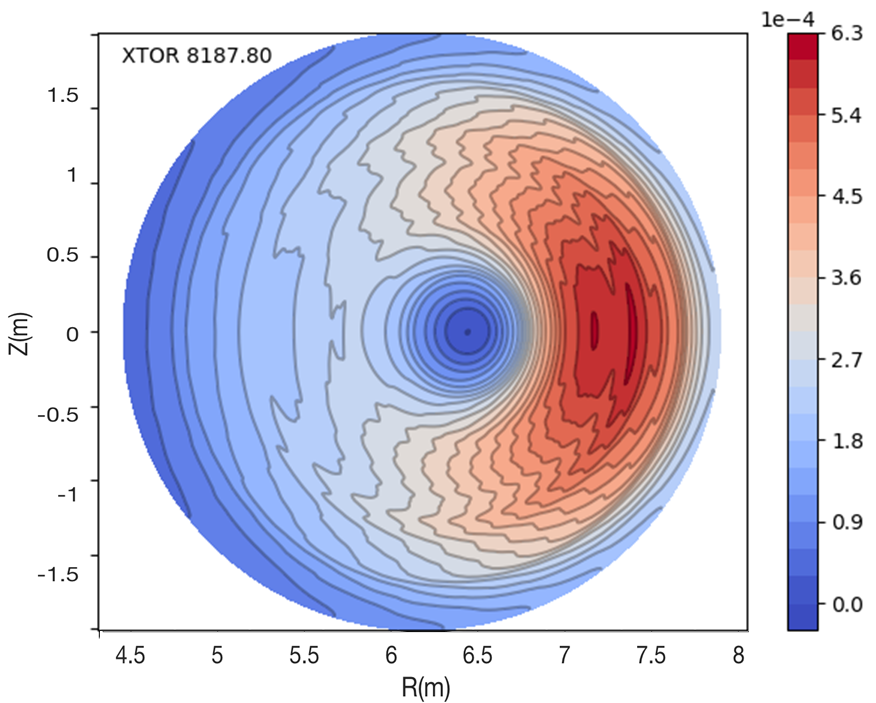}
\caption{Poloidal section of the toroidal velocity $V_\varphi$ (poloidal plane $\varphi=0).$ Velocity is normalized to the Alfv{\'e}n speed.}
\label{ITER_Vphi}
\end{figure}

The other effect of the NB injection is, as expected, the generation of a toroidal torque, due to the tangential injection angle. The toroidal velocity of the bulk, initialized to zero, progressively increases and propagates until it becomes an axisymmetric profile. A poloidal cross-section of the resulting toroidal velocity profile is provided in Fig. \ref{ITER_Vphi}.

\begin{figure}[h!]
\centering
\includegraphics[width=0.5\textwidth]{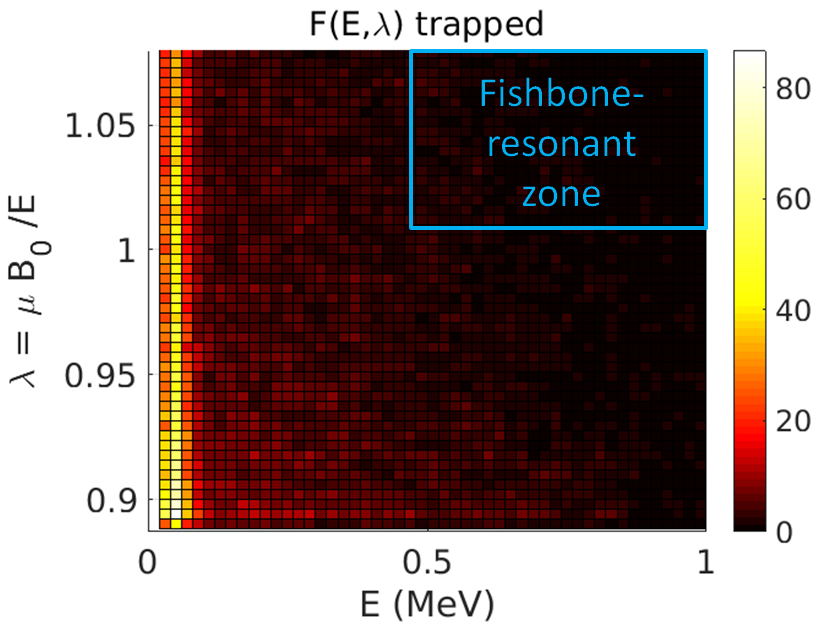}
\caption{Distribution of trapped particles (arbitrary unit) around $q=1$ depending on their energy $E$ and pitch angle $lambda$. The fishbone-resonant region determined in Ref. \cite{Brochard_20} is located inside the light-blue rectangle.}
\label{ITER_Elambda2}
\end{figure}

In a second step, from $t_1=8000~t_A$, injection and collisions are switched off and MHD modes $n=0$ to $11$ are included in the kinetic-MHD simulation. From this time on, an $n=1$ kink mode progressively develops. The same simulation run without NBI and with different NBI power (scanned to scale $\beta_{fp}$ between $0$ and $20\%$) shows that the linear growth rate of the $n=1$ kink mode does not depend on the fast particle pressure. It suggests that no resonant interaction occurs between fast particles and the $n=1$ mode. This statement is supported by linear studies run for similar conditions of fast particles of $1$~MeV in ITER \citep{Brochard_18, Brochard_20}. In this linear study, no resonance is found with passing particles but a resonance with trapped particles destabilize the $n=1$ fishbone mode. Here, as observed in Fig. \ref{ITER_Elambda} (b), most particles are passing: they are characterized by a pitch angle $\lambda<0.88$. In the phase space diagram ($E, \lambda$) presented in Fig. \ref{ITER_Elambda2}, we consider only the minority of particles that are trapped. We notice that most of the trapped particles are thermalized (left of the diagram at low energy). However, almost no trapped particle is present in the fishbone-resonant zone identified in Ref. \cite{Brochard_20} (highlighted in blue in the diagram). Therefore fishbone-resonance with NBI-induced trapped particles appears unlikely in ITER. Note that this statement is also valid with the other injector (NBI \#1) that points towards the tangential direction too. Moreover, according to this simulation in circular geometry, no fishbone resonance with NBI-induced passing particles is observed.   
However, other works in realistic ITER geometry highlights the existence of a resonance with $\alpha$-born passing particles \cite{Brochard_20b}. So future works in realistic geometry will determine whether NBI-induced passing resonance with fishbone can occur in ITER.

\begin{figure}[h!]
\centering
\includegraphics[width=\textwidth]{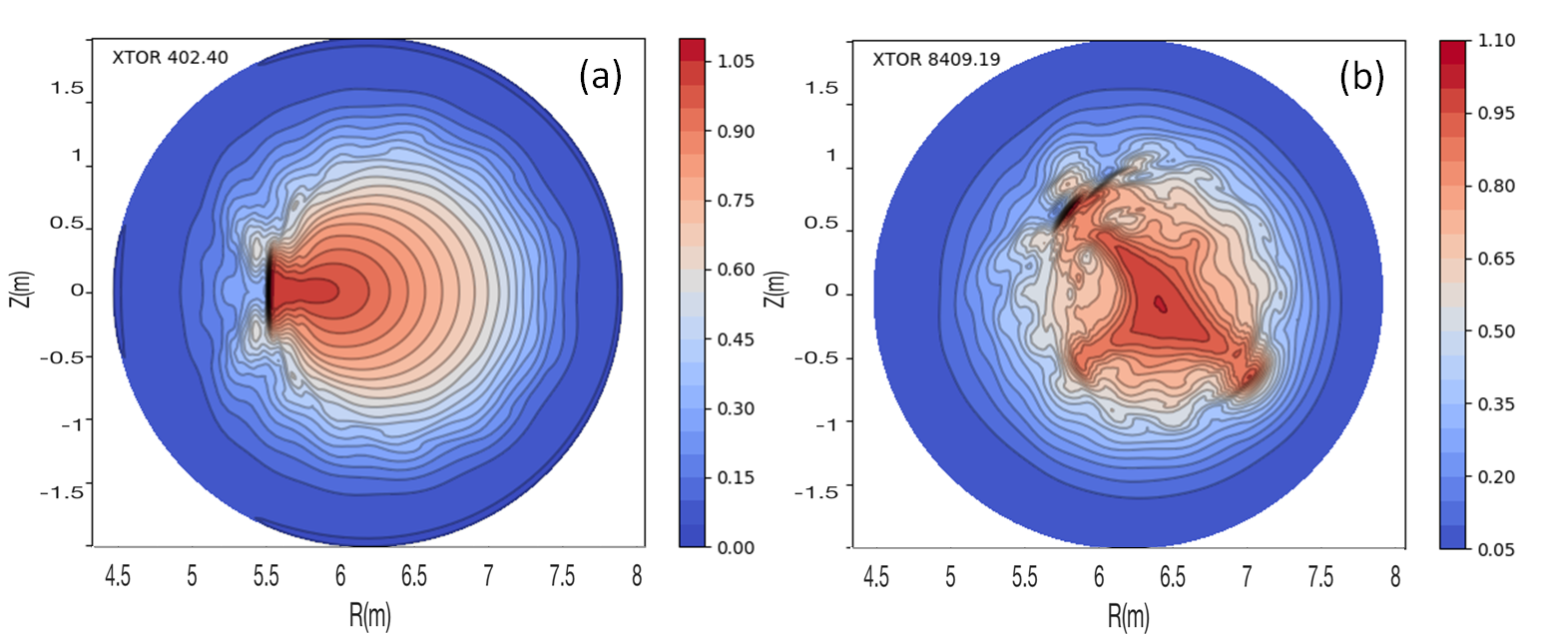}
\caption{Poloidal cross-section of the bulk ion density: (a) without NBI; (b) with NBI. Density is normalized to the initial central density.}
\label{ITER_kink}
\end{figure}

From our simulations in circular geometry, the major impact that NBI-induced fast particles is found to have on the kink stability is the following: they induce an anisotropic total (bulk + fast) pressure profile. As a result, additional harmonics $n=2, 3$ and $4$ non-linearly develop during the kink dynamics. Therefore, even if the time scale of the sawtooth remains the same with or without NBI, the geometry of the mode differs. Fig. \ref{ITER_kink} shows a preliminary result for the kink dynamics in ITER: the bulk ion density is plotted without NBI (left -a) and with NBI (right -b). Without NBI, due to the large pressure in ITER, the $n=1$ kink mode is combined with an interchange mode ($n>8$) inducing finger-like structures in the non-linear phase. With NBI, the same kink-interchange combination is observed but $n=2, 3$ and $4$ harmonics generate three dominant fingers instead of one without NBI. These results are preliminary and should be confirmed in future works.

\section{Conclusion} \label{conclu}

In this paper, a new model of fast ion source induced by Neutral Beam Injection, implemented in the hybrid kinetic-MHD code XTOR-K, has been presented. The injection and ionization model has been designed to reproduce as closely as possible the experimental geometry and parameters. Combined with collisions also recently implemented in XTOR-K, the NBI source allows us to reproduce realistic slowing-down distributions. A limitation is that we need to increase jointly the collision rate and beam power to reduce the computational time. The ionization model tested for an ASDEX Upgrade discharge was shown to qualitatively agree with the results of other NBI models such as NUBEAM. More quantitative comparison will require a benchmark with exactly the same parameters. 

The model has been applied to the interaction of NBI-induced fast ions with a kink mode growing in the plasma core of ASDEX Upgrade. The growth rate of the $n=1$ kink mode was shown to decrease linearly when the NBI power -- and thus the fast particle pressure -- is increased. This partial stabilization is due to the resonant interaction between counter-passing particles with the kink mode, as shown by the accumulation of particles in the resonant region of the ($E,\lambda$) diagram near $q=1$.

A preliminary modeling for an ITER-like case shows that the neutral beam induces a toroidal torque due to the injection of very passing particles. However no resonance was found between NBI-induced fast particles and the kink mode growing in the core. The work of Ref. \cite{Brochard_20} predicts a resonance of trapped particles of energy $\sim 1$~MeV with the fishbone mode. However, almost no trapped particles in the fishbone-resonant range of energy and pitch angle is induced by the beam ionization and slowing-down. That is why no resonance is observed in this ITER-like case. The joint continuation of this study and the work of Ref. \cite{Brochard_20} would be to consider both NBI-induced particles and $\alpha$-born particles cooled down by collisions and consider their interaction with MHD modes.

Even though a simple example of resonant interaction with a kink mode was presented, this new model paves the way to a wide variety of applications in the physics of burning plasmas, which is a hot topic for ITER. In particular, not only the impact of fast particles on sawtooth-fishbone modes, but also the non-linear dynamics of all kinds of fast-particle induced Alfv{\'e}n Eigenmodes could be investigated with this model.



\subsection*{Acknowledgments}

\textsl{This work has been carried out within the ANR project AMICI and within the framework of the EUROfusion Consortium and has received funding from the ANR and the Euratom research and training programme 2014-2018 and 2019-2020 under grant agreement No 633053. The views and opinions expressed herein do not necessarily reflect those of the European Commission. The first author acknowledges G.Tardini for useful input and A.Heron and A.Canou for their support. The first author is also grateful to D.Holmgren and B.Mollison for their inspiring ideas.}

\bibliographystyle{unsrt}
\bibliography{biblio_NBI}

\end{document}